\documentclass[aps,prl,superscriptaddress,twocolumn]{revtex4}

\usepackage {graphicx}
\usepackage {amsmath}
\usepackage {amsfonts}
\usepackage {amssymb}
\usepackage {mathrsfs}
\usepackage {bbm}

\newcommand {\be}{\begin{eqnarray}}
\newcommand {\ee}{\end{eqnarray}}

\newcommand{\ZZ}[2]{\raisebox{#1}[0pt]{#2}}
\newcommand{\fs}{\footnotesize}
\newcommand{\tst}{\textstyle}
\newcommand{\dst}{\displaystyle}
\newcommand{\sst}{\scriptstyle}

\begin{document}

\title {Spin Fluctuation Dynamics and Multiband Superconductivity in Iron Pnictides}
\author {Valentin Stanev}
\affiliation {Department of Physics \& Astronomy, The Johns Hopkins University,
Baltimore, MD 21218}
\author {Jian Kang}
\affiliation {Department of Physics \& Astronomy, The Johns Hopkins University,
Baltimore, MD 21218}
\author {Zlatko Tesanovic}
\affiliation {Department of Physics \& Astronomy, The Johns Hopkins University,
Baltimore, MD 21218}
\date {\today}

\begin{abstract}
Multiband superconductivity, involving resonant pair scattering between different bands, has
emerged as a possible explanation of some of the main characteristics
of the recently discovered iron
pnictides. A key feature  of such interband pairing mechanism is that it can
generate or enhance superconducting pairing {\em irrespective} of whether it is
attractive or repulsive. The latter case typically leads to the superconducting gap
switching its sign among different sections of the Fermi surface.
In iron pnictides, the natural scenario is that the gap changes sign
between the hole and the electron Fermi surfaces. However, the macroscopic symmetry of
such an extended $s'$-wave state still belongs to the general
$s$-wave category, raising the question of how
to distinguish it from an ordinary $s$-wave. In such a quest, it is essential to
use experimental techniques that  have a {\em momentum space} resolution and can
probe momenta of order $(\pi,\pi)$, the wavevector 
that separates the hole and the electron
Fermi surfaces in the Brillouin zone.
Here we study experimental signatures in the spin fluctuation dynamics
of the fully-gapped $s$- and
$s'$-wave  superconducting states, as well as those of the nodal $d$- and $p$-wave.
The coupling between spin
fluctuations of the incipient nearly-nested spin density-wave (SDW) and the Bogoliubov-deGennes
quasiparticles of the superconducting state leads to the Landau-type damping of the former. The
intrinsic structure of the superconducting gap leaves a
distinctive signature in the form of this
damping, allowing it to be used to diagnose
the nature of iron-based superconductivity in neutron
scattering and other experiments sensitive to spin fluctuations
in momentum space. We also discuss
the coexistence between superconductivity and SDW order.

\end{abstract}
\maketitle


\section{Introduction}

Recent discovery of a new high-temperature superconducting family \cite {LaOFeAs, SmOFeAs,
LaOFeAshd,CeOFeAs, PrOFeAs,BaFe2As2,SmFe2As2, SmFe2As2coex,BaFe2As2coex} has generated a flurry of
excitement. Many of the key questions, both theoretical and experimental, remain unanswered.
However, it is becoming rapidly clear that the resemblance to the high-T$_c$ cuprates is less
straightforward and that a new superconducting mechanism might be at play.

Variety of order parameters and pairing mechanisms has been suggested. Phonon interaction alone
seems too weak to explain high $T_c$. In several theoretical papers \cite {Mazin, Vlad, Wang,
Chubukov} the multiband superconductivity in discussed as a possible explanation for the high
$T_c$. Indeed, it has been long known \cite {Suhl} that multiband effects can strongly enhance
superconductivity. This is a particularly relevant in the case of iron-pnictides, which appear to
be moderately correlated electron systems, with
large number of Fe $d$-bands at and near the Fermi
level. Another advantage of this mechanism is that the interband pair interaction can enhance
superconductivity {\em irrespective} of whether it is attractive or repulsive, provided that the
gaps in different bands have the same or opposite signs, respectively. The former is the familiar
$s$-wave while the latter is the extended s-wave
superconductivity, or $s'$. This $s'$ state, with
the superconducting gap having the opposite sign on hole and electron sections of the Fermi
surface (FS), emerges as a natural explanation for the superconductivity in Fe-based compounds,
given the proximity of these materials to various nesting-driven spin density-wave (SDW) and
related instabilities and absence of obvious strong attractive interaction.

The early point-contact Andreev reflection experiments indeed indicated a fully-gapped
superconductor \cite {CLC, Samuely} with no indication of cuprate-like nodes, consistent with this
$s'$ picture, provided that the hole and electron gaps are of a similar magnitude. The subsequent
microwave \cite {Hashimoto} and ARPES \cite {Ding,Hasan} experiments further fortified the case
for $s$ or $s'$ state, also finding the fully-gapped superconductor with, in some instances,
different gaps for the hole and the electron portions of the FS. In contrast, the NMR results are
most naturally interpreted in terms of a $d$- or $p$-wave nodal state \cite{NMR}.

This state of affairs underscores the urgency of settling the issue of the gap structure by
additional experimentation. Very recently, several papers \cite {Scalapino2, Mazin2, Chubukov,
Bernevig} elaborated different possible experimental signatures of the $s'$ superconductivity as
well as of some other forms of superconducting order, particularly in the NMR experiments.
However, in view of the complexity of these materials and still relatively poor quality of the
samples, it is unlikely that a single experiment is going to settle this issue unequivocally.
More importantly, the $s'$ state is in the same symmetry class as the standard $s$-wave and
there is in principle no {\em macroscopic} experiment, analogous to the phase sensitive
measurements in cuprates \cite{Kirtley}, which can distinguish the two in a decisive,
qualitative fashion. Instead, the key difference between the $s'$- and the $s$-wave state
is in the {\em momentum space}, not the real space, and thus one should concentrate
on experiments that have the momentum space resolution and can probe the
wavevectors around $(\pi,\pi)$, which separates the hole ($\Gamma$) and the
electron ($M$) FS in the Brillouin zone.

To this end, and to further expand the range of the experimental techniques that can be used in
this regard, in this paper we consider the spin fluctuations dynamics in the superconducting state
of iron pnictides. We assume that the system is close to a nesting-driven spin density-wave (SDW)
instability and compare different contributions to the damping of spin fluctuations arising from
the Bogoliubov-deGennes (BdG) quasiparticle excitations in superconductors with $s$- and $s'$-wave
gap. For completeness, we also consider the case with nodal $p$ and $d$-wave symmetry of the gap
function. This problem is the superconducting state analogue of the  Landau damping in Fermi
liquids. We find that the damping is qualitatively different in all of the above cases and thus
can be used to diagnose the intrinsic, microscopic nature of the superconducting order. We also
consider the situation in which the ordered SDW/AF state coexists with a superconductor. This
coexistence is observed in at least two of the compounds \cite {SmFe2As2coex, BaFe2As2coex} and
can be induced by pressure or doping.

\section{Preliminaries and The Model}

The parent compound of the 1111 class of Fe-based superconductors has a ZrCuSiAs-type crystal
structure \cite {ZrCuSiAs}, with eight atoms per unit cell. The Fe atoms lie in a plane, same as O
atoms precisely above them, in the adjacent rare earth (RE) oxide layer. In contrast, the RE and
As atoms (also located above each other) are puckered out of plane in a checkerboard fashion. This
puckering of As atoms is crucial for understanding the electronic structure of the compounds
\cite{Vlad}. It brings all Fe d-orbitals close to the Fermi level and creates significant overlap
among Fe and As atomic orbitals. The result is a rich band structure. There are five bands
crossing Fermi surface: two electron cylinders around $M$ point; two hole cylinders plus a hole
pocket around $\Gamma$ point \cite{Mazin, Ma, Lebegue}. The $3D$ hole pocket is quickly filled
with doping and is believed to be irrelevant to both antiferromagnetism and superconductivity. The
remaining two hole and two electron bands are almost 
cylindrical and exhibit significant degree of
nesting. 
The natural Fe magnetism, due to the Hunds rule, is suppressed and one is left with a
weaker itinerant (antiferro)magnetism, sensitive to the nesting features in the band
structure and associated with moderatly strong correlations \cite{Vlad}. 

The outlines of the generic phase
diagram of iron pnictides have began to emerge \cite {OakRidge, Bao1, Bao2}: 
at zero and moderate doping and zero
temperature there is structural distortion and antiferromagnetic order which disappear at some
higher temperature. This critical
temperature is strongly suppressed by doping; at some critical doping structural distortion and
antiferromagnetism suddenly give way to superconductivity. After this point,
further increase in the doping level
produces relatively small changes and the superconducting $T_c$ is rather flat. In addition,
SmO$_{1-x}$F$_{x}$FeAs and Ba$_{1-x}$K$_{x}$Fe$_2$As$_2$ have a sizable region of coexistence of
antiferromagnetism (SDW) and superconductivity.

\begin{figure}[htb]
\includegraphics[scale=1]{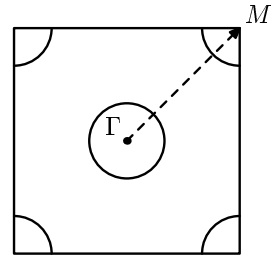}
\caption{An illustration of our model: a hole band (c) centered at the $\Gamma$ point and
an electron band (d) centered at the $M = (\pi,\pi)$ point interact via a short-range
interaction $U$.}
\label{twobands}
\end{figure}

An emerging consensus is that antiferromagnetism is due to the nesting between the hole and the
electron bands. The result is the spin density-wave (SDW) formed  by itinerant electrons. After
particle-hole transformation, the SDW instability is mathematically equivalent to the BCS one,
caused by a logarithmic divergence \cite {Vlad}. Away from perfect nesting this divergence is
replaced with a finite peak and sufficiently far away there is no instability for weak
interactions. Because of proximity to such instability, however, there are enhanced spin fluctuations in
the system. This fluctuations couple to the Fermi liquid quasiparticles and can decay into an
electron-hole pair, which leads to a Landau-type damping. In the antiferromagnetically ordered
state with localized spins or in a perfectly nested SDW  one expects an insulating behavior.
Instead, the experiments show a drop in the resistivity at the SDW transition point followed by
the metallic behavior \cite{resistivity}. This obviously implies 
high degree of itinerancy and indicates that the
Fermi surface is only partially gapped or that the 
Fermi level is located entirely outside the SDW
gap. In presence of the SDW order, the collective excitations of the SDW order parameter (spin
waves) can interact with the quasiparticles on the Fermi surface, thus inducing a Landau-type
damping of the former. Similarly, in the vicinity of the SDW state, when the true long range order
is absent but the correlation length is very large, the spin fluctuations -- the incipient
spin-waves of the SDW -- are also damped by the aforementioned decay into the particle-hole
continuum. When the system undergoes the superconducting transition, this particle-hole continuum
at the Fermi level is gapped by the superconducting order. Consequently, the new fermions  -- the
BdG quasiparticles -- are far less effective in damping the spin fluctuations and the decay rate
vanishes as the temperature goes to zero. This decay rate of spin fluctuations carries a distinct
signature of the structure of the superconducting gap in {\em momentum space} and the associated
BCS coherence factors.

For the purposes of this paper we adopt the following simple yet sufficiently realistic model
depicted in Fig. \ref{twobands}. We assume that the electron spectrum are described by two
different bands: a hole ($c$) band at $\Gamma$ point, and an electron ($d$) band at the $\vec M =
(\pi, \pi)$ point with the same effective mass $m$. This is an approximation to be sure but a
sensible one since the two hole and two electron bands at the FS of real Fe-pnictides resemble
each other to a reasonable degree. In fact, Fe-pnictides are really {\em semimetals}, in the
following sense: consider two bands
\be
\epsilon^c_{\vec p} &=& \varepsilon_c + t_c\cos (p_x a) + t_c\cos(p_y a) \nonumber \\
\epsilon^d_{\vec p} &=& \varepsilon_d + t_d\cos (p_x a) + t_d\cos(p_y a)
\label{semimetal}
\ee
and imagine the situation where $\varepsilon_d -\varepsilon_c\gg t_c,t_d$ and we have
two electrons per unit cell, mimicking the
six d electrons of the Fe-pnictides parent compounds.
The chemical potential is in the gap between the $c$ (full) and $d$ (empty)
bands and the system is
an insulator. Now, as the difference $\varepsilon_d -\varepsilon_c$ is gradually
reduced while the electron number remains unchanged, the bottom of the $d$ band at the
corner $M$ of the Brillouin zone (BZ) moves below the top of the $c$ band at
its center $\Gamma$. The electrons filling the top of $c$ now migrate to the bottom
of $d$ leaving the holes in $c$ behind, thereby
creating the FS shown in Fig. \ref{twobands}.
The density of electrons filling the bottom of $d$ is precisely equal to the
density of holes in $c$ band, hence the semimetal label for the parent compounds. Of course, the
situation in real materials is not as simple as Eq. \eqref{semimetal}; there are
four not two bands and they are far from being simple since their orbital content
changes considerably as one goes around the FS \cite{Vlad}. These complexities
notwithstanding, the above simple picture (Fig. \ref{twobands} and Eq. \eqref{semimetal}) with
$t_c\sim t_d$ will suffice for the purposes of this paper.

The relevant interband
interaction is assumed to be the short-ranged Hubbard $U$.
Thus, the action and the Hamiltonian can be written as

\be
S_0 & = & \int \mathrm{d}\tau\ \mathrm{d}^d r\ \left\{ \bar{c} \left(\partial_{\tau} - \mu \right)
c + \hat{H}^c + \bar{d} \left(\partial_{\tau} - \mu  \right) d + \hat{H}^d \right\} \nonumber \\
S_{int} & =  & U \int \mathrm{d}\tau\ \mathrm{d}^d r\ \bar{c}\; c\; \bar{d}\; d \nonumber \\
\hat{H}^c & = & \sum_{\vec{p}} \epsilon^c_{\vec{p}} \bar{c}_{\vec{p}} c_{\vec{p}} =
\sum_{\vec{p}} \left( \epsilon_F - \frac{\vec{p}^2}{2 m}  \right) \bar{c}_{\vec{p}} c_{\vec{p}} \nonumber \\
\hat{H}^d & = & \sum_{\vec{p}} \epsilon^d_{\vec{p}} \bar{d}_{\vec{p}} d_{\vec{p}} = \sum_{\vec{p}}
\left(- \epsilon_F + \frac{(\vec{p} - \vec{M}) ^2}{2 m}  \right) \bar{d}_{\vec{p}} d_{\vec{p}}
\label{hamiltonian}
\ee
where we have expanded $c~(d)$ band \eqref{semimetal} near the top (bottom). We can now shift the
electron band to the $\Gamma$ point, and call this "new" electron band $e$. One should, however,
always keep in mind that the two bands are shifted relative to each other in momentum space by
$\vec M = (\pi, \pi)$ (Fig. \ref{twobands}).


First, we consider the case of perfect nesting $\mu = 0$ 
and compute the damping term for the spin fluctuations, given by the imaginary part of the
electronic spin susceptibility:
\begin{eqnarray*}
\chi (\vec{q} + \vec{M}, \omega) = \sum_{\vec{p}} \frac{ f(\epsilon^c_{\vec{p} + \vec{q}}) -
f(\epsilon^e_{\vec{p}})}{\omega - (\epsilon^c_{\vec{p}+ \vec{q}} - \epsilon^e_{\vec{p}}) + i 0^+}
\end{eqnarray*}
which results in:
\begin{equation}
Im \chi (\vec{q} + \vec{M}, \omega) \propto \frac{\omega}{v_F q}~,
\end{equation}
having the familiar Landau damping form.

\begin{figure}[htb]
\includegraphics{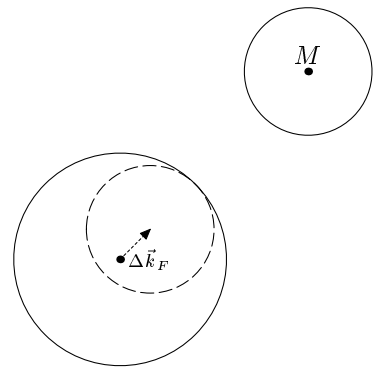}
\caption{Fermi surfaces of inequivalent hole and electron bands. Their centers are
displaced by the vector $\vec {\Delta k_F} $ defined in the text.}
\label{differentFS}
\end{figure}

Next, we consider a more realistic case and assume  that two bands are mismatched - their centers
do not coincide -- one being  shifted by $\vec {\Delta k_F} $ -- and furthermore their radii
differ by $2 \mu$, as illustrated in Fig. \ref{differentFS}. 
These two quantities, $\vec {\Delta
k_F} $ and $\mu$, parameterize deviations from perfect 
nesting within our model. The spectrum of
the two bands is:
\begin{eqnarray*}
\epsilon^c_p & = & \epsilon_F - \frac{\vec p^2}{2 m} - \mu \\
\epsilon^e_p & = & - \epsilon_F + \frac{(\vec{p} + \vec {\Delta k_F})^2}{2 m} - \mu~.
\end{eqnarray*}
Notice that the damping term is nonzero only if the two Fermi surfaces touch or intersect each
other, otherwise it is impossible  to excite an electron-hole pair with a low energy
fluctuation and momentum close to $\vec{M}$. When two Fermi surfaces have two different
intersection points, the damping term is proportional to $\omega$: 
\be
 \frac{ \omega }{\sqrt{ \left( \frac{P_0 \vert\Delta k_F\vert} {2 m} \right)^2 - \mu^2}}~,
\ee
where  $P_0^2 = 2 m \epsilon_F - (\vert\Delta k_F\vert /2)^2$.

When the two bands are touching each other at a point on the Fermi surface, $\vert\Delta k_F\vert
= \sqrt {2 m (\epsilon_F + \mu)} - \sqrt {2 m (\epsilon_F - \mu)}$.
For $\mu \ll \epsilon_F$ but still finite, the low frequency damping term is:
\be
Im \chi (\vec{q} + \vec{M}, \omega) \propto \frac{\omega}{2 \mu}~.
\ee

\section{Spin Fluctuations in a Superconducting State}

Once our system enters the superconducting state,  the damping changes and becomes temperature
dependent reflecting the opening of the superconducting gap.
The coupling of the spin fluctuations to the new
excitations -- the BdG quasiparticles -- is determined by the BCS coherence factors \cite
{Tinkham}. These coherence factors, however, have a  different form for different forms
of the {\em microscopic}
superconducting order parameter -- the BdG gap function -- and
therein lies the possibility for diagnosing the
intrinsic nature of the superconducting state.

Again, we first consider the case of perfect nesting between the hole and the electron band(s).
Since we are interested in scattering of BdG quasiparticles and in the processes that involve
spin, the correct coherence factor to use is $u_{\vec{p}} u_{\vec{p} + \vec{q}} + v_{\vec{p}}
v_{\vec{p} + \vec{q}}$ or, explicitly:
\be
  \frac {1}{2} \left(1 + \frac {\epsilon^c_{\vec{p}} \epsilon^e_{\vec{p} + \vec{q}} + \Delta^c
  \Delta^e }{ E^c_{\vec{p}} E^e_{\vec{p}+ \vec{q}}} \right)~,
\ee
where $E^c_{\vec{p}} = \sqrt {(\epsilon^c_{\vec{p}})^2 + (\Delta^c)^2}$  and $E^e_{\vec{p}} =
\sqrt {(\epsilon^e_{\vec{p}})^2 + (\Delta^e)^2}$. 
Since our focus in on spin fluctuations with
momenta around $ \vec {M} = (\pi, \pi)$, we are 
justified in dropping the intraband terms.
The electron susceptibility then can be written as
\be
\lefteqn{\chi (\vec{q} + \vec{M}, \omega) = }   \\
& = & \sum_{\vec{p}} \frac {1}{2}\left(1 + \frac {\epsilon^c_{\vec{p}} \epsilon^e_{\vec{p} +
\vec{q}} + \Delta^c_{\vec{p}} \Delta^e_{\vec{p}+\vec{q}}}{E^c_{\vec{p}} E^e_{\vec{p} +
\vec{q}}}\right) \frac {f(E^c_{\vec{p} + \vec{q}}) - f(E^e_{\vec{p}})}{\omega - (E^c_{\vec{p} +
\vec{q}} - E^e_{\vec{p}}) + i 0^+} \nonumber \ee
where $ q \ll M$. Extracting the imaginary part
gives the general expression for damping of spin fluctuations at this momentum
by the BdG particle-hole excitations:
\be
\lefteqn{Im \chi (\vec{M} + \vec{q}, \omega)  = } \nonumber \\
& = & \sum_{\vec{p}} \frac {1}{2} \left(1 + \frac {\epsilon^c_{\vec{p}} \epsilon^e_{\vec{p} +
\vec{q}} + \Delta^c_{\vec{p}} \Delta^e_{\vec{p} + \vec{q}}} {E^c_{\vec{p}} E^e_{\vec{p} + \vec{q}
}} \right)  \nonumber \\
& & \times \bigl(f(E_{\vec{p} + \vec{q}}) - f(E_{\vec{p}})\bigr) \delta\bigl(\omega - (E_{\vec{p} + \vec{q}} -
E_{\vec{p}} ) \bigr)~.
\label{delta}
\ee

First we consider the $s'$ case. Because of  $\Delta^c_{\vec{p}} = - \Delta^e_{\vec{p} +
\vec{q}}$ ($s'$ implies the repulsive pairing term) the coherence factor
is strongly suppressed $E_{\vec{p}}
E_{\vec{p}+ \vec{q}} + \epsilon^c_{\vec{p}} \epsilon^e_{\vec{p} + \vec{q}} - \Delta_{\vec{p}}
\Delta_{\vec{p} + \vec{q}} \approx \frac{1}{2} ((v_F q \cos\theta)^2 (1 + (\frac{\epsilon_{\vec{p}}}
{E_{\vec{p}}})^2)$, after expansion in the powers in $q$ (we dropped the band indices for the moment and put $\epsilon^c_{\vec{p}} \equiv \epsilon_{\vec{p}}$). The argument of
the $\delta$-function in \eqref{delta} becomes
(using $E_{\vec{p} + \vec{q}} - E_{\vec{p}} \approx \frac {\epsilon_{\vec{p}} v_F q
\cos\theta}{E_{\vec{p}}}$ for small $q$) $\approx \omega - \frac{\epsilon_{\vec{p}} v_F q
\cos\theta}{E_{\vec{p}}}$, with $\cos\theta$ having
to be positive for $\epsilon^c_{\vec{p}}$. We can now
use $\epsilon_{\vec{p}} = \sqrt {E_{\vec{p}}^2 - \Delta_{\vec{p}}^2}$ to solve for $E_{\vec p} \equiv E$:
\be
\lefteqn{\delta \bigl(\omega - (E_{\vec p + \vec q} - E_{ \vec p})\bigr) =} \nonumber \\
& = &  \frac{1} {v_F q \cos\theta} \frac {E^2 \sqrt {E^2 - \Delta^2} }{\Delta^2} \delta \left(E -
\frac {\Delta}{\sqrt{1 - x^2}} \right)~, \nonumber
\ee
where $x = \frac {\omega}{v_F q \cos\theta}$ is obviously smaller then unity; we
will assume the condition $\frac {\omega} {v_F q} \ll 1$, as usual.
The occupation number factor in Eq. \eqref{delta}, $ f(E_{\vec p + \vec q})
- f(E_{ \vec p})$, can be written as
$f(E_{\vec p} + \omega) - f(E_{\vec p}) \approx \omega \frac
{\partial f(E)}{\partial E}$. We then convert the $p$ integration into $E$ integration, with
density of states $ \frac {E}{\sqrt {E^2 - \Delta^2}}$, and
perform the integral over the $\delta$-function. This leaves us with
only the angular integration remaining:
\be
\int_{-\frac{\pi}{2} + \frac {\omega} {v_F q}}^{\frac{\pi}{2} - \frac {\omega} {v_F q}}  (v_F q \cos\theta) \left(- \frac {2E^2 - \Delta^2 }{E \Delta^2}\right)(\omega \frac {\partial f(E)}{\partial E}) dE \nonumber \\
= \int_{-\frac{\pi}{2} + \frac {\omega} {v_F q}}^{\frac{\pi}{2} - \frac {\omega} {v_F q}} (v_F q \cos\theta)  \frac {(1 + x^2)}{ \sqrt{1- x^2} \Delta} \omega \frac {\partial f(E)}{\partial E} dE ~,\nonumber \\
\ee
where $\frac {\partial f(E)}{\partial E}$ is evaluated at $ E = \frac {\Delta}{\sqrt{1 - x^2}}$.
This integral is non-trivial and to perform
 it we have to use $\omega/v_F q \ll 1$. Despite the $\sqrt{1- x^2}$
in the denominator the integrand is rapidly
suppressed in the limit $x \rightarrow 1$, because of
the  exponential factor hiding in $\frac {\partial f(E)}{\partial E}$ .
It is straightforward  to check that the
first derivative at $\theta = 0$ vanishes and that
 the integrand has a maximum there. Because of the
rapid decrease away from this point, we can estimate the
integral with the familiar formula $I \approx y(0)
\sqrt {\frac {y(0)} {\vert y^{''}(0)\vert}}$, where $y(\cos\theta)$ is the integrand.
Working to the lowest order in $\frac {\omega} {v_F q}$ we finally obtain:
\be
 Im \chi^{s'} (\vec q, \omega)  \propto  \omega (v_F q)  \frac {e^\frac {\Delta}{T}}{(e^{\frac {\Delta}{T}} + 1)^2 T}~.
\ee
This expression gives the interband contribution to
the damping of an incipient fluctuating spin wave in an $s'$-wave superconductor.

It is instructive to contrast the above result with the standard $s$-wave case. After performing
the analogous calculation for the case of a pure $s$-wave superconductor,
the coherence factor for
the superconductor is found to have a constant term $\sim \frac {\Delta^2}{E^2}$ . Keeping only
this term and going trough the same steps as before yields:
\be
Im \chi^s (\vec q, \omega) \propto
\frac{\omega} {v_F q}  \frac {e^\frac {\Delta}{T}}{(e^{\frac {\Delta}{T}} + 1)^2 T}~.
\ee
Clearly, the dynamical properties of spin fluctuations in $s$ and $s'$
superconducting phase are very
different: the damping and thus the decay rate of incipient spin waves in the $s'$ state is
substantially reduced relative to the standard $s$-wave case,
by the overall factor of $q^2$ (note
that ${\vec q}$ is measured {\em relative} to ${\vec M}$). These differences and the specific
forms of damping should be observable in neutron scattering or other momentum-resolved probes of
spin fluctuations.

For completeness, we now consider the nodal $d$- and $p$-wave superconductors, another
possible contenders for the superconducting state of iron pnictides, favored by
the NMR experiments \cite{NMR}.  In this case, the main contribution to the low temperature
damping comes from the nodal regions. The natural excitations of the system in this regions are
Dirac fermions with linear dispersion $v_F \vert p\vert$, which can be seen by expanding the
BdG Hamiltonian near the nodes \cite {Simon}. We will ignore the
intrinsic anisotropy of the BdG-Dirac spectrum, since we do not
expect it is to change the overall form of the damping term.
The coherence factors for the individual nodes differ for
the intra and interband processes, but we need to add all the nodal contributions. The combined coherence factors
 are equal to lowest order (up to a numerical prefactor). First, we assume the $p_x p_y$
$d$-wave order parameter ($d_{xy}$), which means that the direction of $\vec{M}$
and the nodal directions are rotated by $n \pi/4$
(where $n$ is integer) with respect to each other. We expand around that nodes
and do the  momentum  integral first, using the $\delta$-function:
 \be
 \omega \int \delta(\omega - E_{p+q} + E_p) \frac
{\partial f(\epsilon)}{\partial \epsilon} E dE, \nonumber
 \ee
where now $E = v_F\sqrt {p_x^2 + p_y^2}$ is the linear dispersion of the Dirac quasiparticles. The integrand in the remaining angular integral is confined in the region $\theta \in
(\pi/2, 3\pi/2)$, and is peaked around $\theta = \pi$. We again use the formula $I \approx y(0)
\sqrt {\frac {y(0)} {\vert y^{''}(0)\vert}}$ to estimate the lowest order contribution. This gives
the final low temperature limit result: \be
 Im \chi^{d_{xy}} (\vec q, \omega) \propto \omega \sqrt{q^2_x + q^2_y}  \frac{1}{T \cosh^2{\frac {\vert q\vert}{T}}}.
\ee
The reason for this particular temperature dependence is the fact that we
have restricted ourselves to the region $\omega \ll v_F q$.
Integrating this expression over $q$ we can obtain the standard result
for the NMR damping rate $1/(T_1 T)\propto T^2$.

The other possibility for a $d$-wave order parameter is a $d_{x^2 - y^2}$. In this case the nodal
directions are along $\vec {M}$ or perpendicular to it. To the lowest order the damping in this
case has the same form as in the $d_{xy}$ case, so the position of the nodes with respect to $\vec
M$ is irrelevant. We emphasize that this is only the lowest order result, so in general some
distinction between the $d_{xy}$ and $d_{x^2 - y^2}$ cases is expected.

 Now let us consider a $p$-wave superconductor with
two nodes along $x$ direction. Combining the coherence factors for the nodes gives an additional
factor of $\sin^2{\theta}$, compared to the $d_{xy}$ case. We again use the $\delta$-function to
do the momentum integral first. After that the integrand is confined in the $(\pi/2, 3\pi/2)$
interval, but now goes to zero at $\theta = \pi$. Nonetheless, we can still estimate the integral
and the leading term behavior is the same as in the $d_{xy}$ case (with a different numerical
prefactor):
  \be
 Im \chi^{p_{x}} (\vec q, \omega) \propto \omega \sqrt{q^2_x + q^2_y}  \frac{1}{T \cosh^2{\frac {\vert q\vert}{T}}}.
\ee
  We obtain the same result for a $p$-wave superconductor
with nodes along the $\vec M$ direction.

A multigap nodal superconductor can also be an extended $d$ and $p$-wave, with sign change between
the gaps on the disconnected parts of the Fermi surface: a $d'$ or $p'$ state. The lowest order
term in the coherence factor is then proportional to $q^2$, as in the case of $s'$ superconductor,
and the damping in $d'$ and $p'$ states is strongly suppressed compared to the pure $d$ and
$p$-waves. We see that momentum resolved measurements can distinguish the $p$ and $d$-wave
superconductor from the $s$ and $s'$ case. This comes in addition to the significant differences
in their temperature behavior.

Perfectly nested bands generate strong spin and
charge density-wave (SDW and CDW) instabilities.
It is believed that superconductivity appears with
doping or by application of pressure, away from
the perfect nesting, once the ordering in the particle-hole channel is suppressed. Thus, we now
turn to the case of imperfect nesting. To this end, we again consider the two bands -- a hole and an
electron one, illustrated in Fig. \ref{differentFS}. We also assume that the deviation from perfect
nesting is small and $\Delta k_F$ is of the order of $q$ and smaller that $\Delta /v_F$. Now
expanding the coherence factor in the $s'$ case gives us  $  \frac{1}{2}  ( 2 \mu + \vec{p} \cdot
(\vec{q} + \Delta \vec k_F )/m)^2 (1 + (\frac{\epsilon_p} {E_p})^2)$. Repeating the same steps as
before we obtain the damping: \be
 Im \chi^{s'} (\vec q, \omega) \propto  \omega ( 2 \mu + v_F \vert \vec {q} + \vec {\Delta k_F} \vert ) \times \nonumber \\
 \times \sqrt {\frac { 2 \mu + v_F \vert \vec {q} + \vec {\Delta k_F} \vert  }{v_F \vert \vec {q} + \vec {\Delta k_F} \vert }}  \frac {e^\frac {\Delta}{T}}{(e^{\frac {\Delta}{T}} + 1)^2 T}.
\ee
For the $s$-wave superconductor coherence factor we again keep only $\Delta^2$ term and the result is
\be
 Im \chi^s (\vec q, \omega) \propto \frac { \omega} { 2 \mu + v_F \vert \vec {q} + \vec {\Delta k_F} \vert  } \times \nonumber \\
 \times \sqrt {\frac { 2 \mu + v_F \vert \vec {q} + \vec {\Delta k_F} \vert  }{v_F \vert \vec {q} + \vec {\Delta k_F} \vert }}  \frac {e^\frac {\Delta}{T}}{(e^{\frac {\Delta}{T}} + 1)^2 T}.
\ee

Both results reduce correctly to the perfect nesting case ($\mu \rightarrow 0, \Delta k_F
\rightarrow 0$) but generally have more complicated behavior. Increasing the mismatch of the two
bands drives the damping  to the same form  $const \times \omega$ in both cases, albeit with
different constants.

  In the case of $d$- and $p$-wave superconductors, the imperfect nesting of this type leads to a rather complicated behavior. The point where the two Fermi
surfaces touch (Fig. \ref{differentFS})
is special and its position with respect to the nodes determines
the low-energy, low-temperature response. If there is a node in a
close vicinity of this point of contact between two Fermi surfaces, the behavior of the system
is somewhat similar to a perfectly nested single node
case, but with $\vec M$ replaced by $\vec M + \vec {\Delta k_F}$.
If this point is in one of the anti-nodal
regions the response resembles the one found in the $s$-wave case.

We pause here to emphasize again that
the expressions for damping of spin fluctuations presented above -- and similar other physical
quantities -- at a {\em specific} momentum around $\vec {M}$ are  better
suited to distinguish between different {\em microscopic} order parameters and
their associate BdG gap functions. In contrast, the NMR relaxation rate is
an integral quantity and thus always includes {\em both} the interband {\em and}
the intraband contributions. Extracting one of them
is thus a rather challenging task. For example, even in the
$s'$ case, the  Hebel-Slichter peak is expected to occur
\cite{Bernevig}, due to the intraband (pure $s$) processes.

\section{Coexistence of Superconductivity and SDW}

\begin{figure}[h]
\scalebox{2.0}[1.5]{\includegraphics{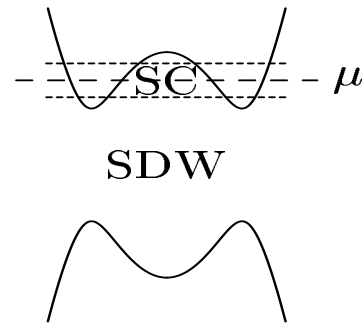}}
\caption{Coexistence of spin density
wave (SDW) and superconductivity in our model. The chemical potential $\mu$ is above the SDW gap
while the superconducting gap is always pinned to $\mu$.} \label{coexistence}
\end{figure}

We now consider the combination of SDW and superconducting orders. 
We will assume that the main SDW
(particle-hole) gap opens below the Fermi level, 
so the system remains metallic, as it appears 
to be the case experimentally \cite{resistivity}.
This is depicted in Fig. \ref{coexistence} and it emulates
the situation in real Fe-pnictides, where the 
mismatch among hole and electron pockets results in reconstruction
instead of complete disappearance  of the Fermi surface \cite{Vlad,vdw}. 
Then -- on top of this SDW -- either attractive or 
sufficiently strongly repulsive interband
interaction generates at some lower temperature a 
new {\em superconducting} (particle-particle)
gap, this time located precisely on the Fermi 
surface (see Fig. \ref{coexistence}).  The question
is whether spin waves  dynamics can be used as a 
probe of the {\em microscopic} structure of the
superconducting order parameter in this case as well.

The magnon (spin-wave) dispersion relation is given by the
inverse of the transverse magnetic susceptibility
$\chi^{+-}$ and is  $\omega^2 - c^2 q^2 $ at the bare level, where $c$ is
the spin-wave velocity.
Damping will be introduced by including
magnon-electron interactions, as before. Here we neglect
contributions from impurities and higher order
magnon-magnon interactions; these are generally present but can be distinguished
from the itinerant particle-hole damping by their different and lesser sensitivity
to the opening of the superconducting gap. Then the imaginary part
of the electron spin susceptibility $\chi$
determines the decay rate of magnons. In the SDW-only phase we have to rewrite the electron
operators as a superposition of the new eigenstates,
which diagonalize the broken symmetry Hamiltonian. This leads
to the appearance of new coherence factors, reflecting the SDW
order in the particle-hole channel \cite {Gruner}. Since we are interested in scattering
of quasiparticles and in processes that involve spin, the correct coherence
factor to use is $u_{\vec p}
u_{\vec p + \vec q} - v_{\vec p} v_{\vec p + \vec q}$:
\be
  \frac12 \left(1 - \frac {\epsilon_{\vec p} \epsilon_{\vec p + \vec q} + \phi_0^2}{\xi_{\vec p} \xi_{\vec p+ \vec q}} \right)
\ee
where $\epsilon_{\vec p} = p^2/2m - \epsilon_F$ and $\xi_{\vec p} = \sqrt {\epsilon_{\vec p}^2 +
\phi_{0}^2}$, with $\phi_0$ being the SDW order parameter.

In presence of the SDW, the unit cell doubles and $\vec{M}$ becomes a vector of the
reciprocal lattice. We can picture the new band structure with two bands at the same position in
momentum space, with avoided crossing where the SDW gap opens. What was previously a hole-like
(electron-like) band now has an electron-like (hole-like) part as well.

One important point should be discussed here. To have metallic behavior and superconductivity on
the top of SDW we have to introduce the non-zero chemical potential for the new (gapped) band
structure. This, however, cannot be done cavalierly, since the model will not be
self-consistent. That is the consequence of the fact that within the simple model
adopted in this paper, a gap on
the Fermi surface is typically
energetically preferable to a gap below or above. To remedy this and
stabilize the SDW gap below the
Fermi surface, we have to include new terms in our Hamiltonian, reflecting 
deviations from idealized bands and 
the lattice effects in real iron pnictides. This, unfortunately, comes at the cost of
significantly more complicated calculations and an entirely obscured physical picture.
For the purposes of this paper, this is too great a cost 
and we avoid it by using the
following approach: we calculate the SDW wave coherence 
factors at zero chemical potential, when
Fermi surface is completely gapped.
Next, in order to account for the metallic behavior, we reintroduce a small Fermi
surface, which now can undergo a superconducting instability.
The physics behind this approach is clear: the SDW coherence factors
determine the vertices that couple the magnons and the BdG quasiparticles.
The magnons are the Goldstone modes of the SDW
and this coupling should be of a gradient type, because of the broken
symmetry of the ordered state -- the long wavelength twist in the spin direction of the SDW
should come at no cost in energy.
This remains true independently of the position of the SDW gap.
Adopting this simplification, the SDW coherence factor is just $\sim (v_F q \cos\theta)^2$.

Starting with the case without superconductivity, the damping is:
\be
 Im \chi (\vec q, \omega) \propto \frac{\omega^3}{v_F q} \frac{\mu \phi_0^2}{4 (\mu^2 -
 \phi_0^2)^2}~.
\ee
  In the case of coexistence we have to calculate anew the
{\em superconducting} coherence factor
\be
  \frac {1}{2} \left(1 + \frac {\xi_{\vec p} \xi_{\vec p + \vec q} + \Delta_{\vec p} \Delta_{\vec p + \vec q}}{E_{\vec p} E_{\vec p + \vec q}} \right)~.
\ee
In the vicinity of the SDW gap, the new quasiparticles are almost an
equal mixture of the two initial bands, $c$ and $d$. Turning on
the chemical potential moves the Fermi level to the upper band;
as discussed above this is to be done advisedly.
If the chemical potential is not very small, the content of the
quasiparticles at the new Fermi level will be almost exclusively from
only one of the old bands (either purely an electron-like or
a hole-like). With this assumption we
calculate the superconducting coherence factors.

Again, we first consider the $s'$ case. Working to lowest order in $\omega /q$ we get:
\be
 Im \chi (\vec q, \omega) \approx \omega (v_F q)^3  \frac {e^\frac {\Delta}{T}}{(e^{\frac {\Delta}{T}} + 1)^2 T}.
\ee This expression gives the interband contribution to the spin wave damping in an $s'$-wave
superconductor. However, here we cannot neglect the intraband contribution
as we did earlier, since  now the momenta $\vec{M}$  and $0$ are effectively
equivalent by virtue of the umklapp scattering off the underlying
SDW modulation -- consequently
{\em both} inter and intraband terms are contributing.
Calculating the intraband term gives:
\be
 Im \chi (\vec q, \omega) \approx \omega (v_F q)  \frac {e^\frac {\Delta}{T}}{(e^{\frac {\Delta}{T}} + 1)^2 T}.
\ee
and we see that the this is in fact the leading term in the limit $q \rightarrow 0$,
masking the contribution of the interband damping.

For the case of a pure $s$-wave superconductor we get
\be
 Im \chi (\vec q, \omega) \approx \omega (v_F q)  \frac {e^\frac {\Delta}{T}}{(e^{\frac {\Delta}{T}} + 1)^2 T}.
\ee
The intraband term is of the same order.


\begin{table}[h]
\begin{tabular}{c||c|c|c|c|c}
       &{\scriptsize pure} & \ZZ{-1.6ex}{\fs $\tst s'$ SC} & \ZZ{-1.6ex}{$\dst s$} & \ZZ{-1.6ex}{$\dst d_{x^2 - y^2}$ } & \ZZ{-1.6ex}{$\dst p_x$} \vspace{-1.6ex}\\
       &{\tiny SDW}  &         &         &         &           \\
\hline \hline
 \ZZ{-1ex}{\scriptsize perfect} &  \ZZ{-2ex}{$\dst \frac{\omega^3}{q}$}      & \ZZ{-2ex}{$\dst \omega q e^{-\frac{\Delta}{T}}$}  & \ZZ{-2ex}{$\tst \frac{\omega}{q} e^{-\frac{\Delta}{T}} $}
     & \ZZ{-2ex}{$\tst \frac{\omega \vert q \vert}{T \cosh^2 {\frac {\vert q\vert}{T}}} $}    &   \ZZ{-2ex}{$\tst \frac{\omega \vert q \vert}{T \cosh^2 {\frac {\vert q\vert}{T}}} $}  \vspace{-2.5ex}    \\
 \ZZ{1ex}{\scriptsize nesting} &         &          &         &        &         \\ \hline
 \ZZ{-1ex}{\scriptsize imperfect} & \ZZ{-2.2ex}{$\dst \omega$}   &  $\sst (2\mu+\alpha_{\vec q}) e^{-\frac{\Delta}{T}} \times $    &
 $\sst \frac{\omega}{(2\mu+\alpha_{\vec q})} e^{-\frac{\Delta}{T}} \times $    &        &   \\
 \ZZ{1ex}{\scriptsize nesting} &         &     $\times \omega \sqrt {\frac { 2\mu+\alpha_{\vec q} }{\alpha_{\vec q}}} $  &    $\times \sqrt {\frac { 2\mu + \alpha_{\vec q} }{\alpha_{\vec q}}} $     &        &
\end{tabular}
\caption{The summary of our results for damping of spin fluctuations by particle-hole excitations
in several different superconducting states. Here $\alpha_{\vec q}= v_F\vert\vec {q} + \vec{\Delta
k_F}\vert$ .}
\label{tbl:damp2} 
\end{table}

Thus, in the region of coexistence, the dynamical properties of SDW magnons in an $s$ and $s'$
superconductor are difficult to distinguish. The damping at momentum around $\vec{M}$ is due to
both inter and intraband scattering and since the latter are, of course, unaffected by the
relative sign of the different gaps, it would be difficult to observe any significant difference
in experiments.

Table \ref{tbl:damp2} summarizes some of our results for the reader's convenience.

\section{A simple model of Multiband superconductivity}

A significant part of our focus in this paper was on an $s'$-wave superconductor, in
which the Fermi surface of Fig. \ref{twobands} is fully gapped 
but the gap function in its hole pocket
has the opposite sign relative to the one in the electron pocket.
Furthermore, our paper is clearly designed to arouse interest in the experimental
community. Consequently, for the reader's benefit, we consider here a rather
basic picture of $s'$ multiband superconductivity and some of its
features, appropriate for our simplified model
of Fe-pnictides (for a more theoretically inclined discussion, the reader is referred to
\cite{vdw}). With appropriate modifications, the same considerations can be easily
adapted to the $d'$ and $p'$ cases.

First, the model \eqref{hamiltonian} discussed so far is not sufficient. The interband
interaction $U\bar c c\bar dd$ will induce the SDW and promote strong spin fluctuations
but will not by itself lead to superconductivity from purely electronic interactions
\cite{vdw}. For that possibility to enter into play, one must consider
the interband pair resonance interactions of the type $J\bar c\bar c dd + h.c.$,
a Josephson-like term in the momentum space which scatters pairs of electrons
between $c$ and $d$ bands \cite{vdw,Chubukov}. Such terms are typically
present in multiband systems and their size in Fe-pnictides is significant \cite{vdw}.

The key feature of this interband interaction $J$ is that it can drive the
system superconducting -- or enhance the already present intraband superconductivity --
{\em irrespective} of whether it is attractive or repulsive. In the former case,
the intraband gaps will have the same relative sign while in the latter, which
is probably where the Fe-pnictides belong, this sign will be different. This is
the origin of the $s'$ superconducting state.

In the weak-coupling  theory,  however, the interband repulsive interaction $J$ can drive the
system superconducting {\em only} if it is large enough, i.e., $J^2 > U_1 U_2$, where $U_1$ and
$U_2$ are the {\em repulsive} intraband interactions in the hole ($c$) and electron ($d$) bands.
Such a sizeable $J$ is not something that is easily found, since, generically, $J$ is
significantly smaller than the intraband Coulomb repulsion $U_1$, $U_2$; indeed, this also appears
to be the case in Fe-pnictides \cite{vdw}. Here 
we suggest a mechanism that could potentially
solve this problem. The interactions that enter the condition 
for superconductivity are not the
bare Coulomb terms but some appropriately screened interactions. The screening length is $
\kappa_D = 4 \pi e^2 \Pi_0 (q) $, where $ \Pi_0 (q) $ is a polarization bubble. The momentum
region for the $J$ pairing term is around $M$ and the main contribution to the polarization is a
mixed bubble. On the other hand the main region for the $U_{1,2}$ terms is around $0$ and the
screening is due mostly to the usual single-band bubbles. $ \kappa_D (\vec M)  $ has a rather
dramatic evolution with doping \cite{Vlad} (in contrast with  $ \kappa_D (0) $), and goes to $0$
for high doping levels. Let's now confine  ourselves to the region in which the interactions are
frequency independent. This limit is constrained by the time $\tau$ it takes an electron to
traverse the Debye screening length $\tau \approx \frac{r_D}{v_F}$ where $r_D= \frac{1}{\kappa_D}
= \frac{1}{4 \pi e^2 \Pi_0}$. The interaction is frequency independent for $\omega \ll
\frac{1}{\tau}$. Outside this region it has complicated frequency dependence and we can set $J=0$
(as it is done in the extensions of the BCS theory which include Coulomb interaction, where $J =
const$ for $\omega < \omega_C$ and $J=0$ for $\omega > \omega_C $, where $\omega_C$ is the Coulomb
frequency cut-off). For $ \kappa_D (0) > \kappa_D (\vec M) $, the frequency range of $J$ is {\em
smaller} and this could lead to logarithmic suppression of $U_{1,2}$ 
akin to the reduction in the
Coulomb pseudopotential caused by phonon retardation 
effects within the standard BCS-Eliashberg theory \cite{Littlewood}.

We now illustrate the above arguments with an explicit calculation.
In multiband superconductivity there are at least two gaps which vanish at the same critical
temperature (if there is a non-vanishing pair resonance term $J$). 
We can write down the equations
determining the zero temperature gaps and the 
critical temperature \cite {Suhl,Kondo}. To study
the effects of the retardation we divide the energy range 
into two intervals: $\epsilon \in (0,
\omega_{C1})$ and $\epsilon \in (\omega_{C1}, \omega_{C2})$, 
where $\omega_{C1}$ and $\omega_{C2}$
are the frequency cut-offs for $J$ and $U_{1,2}$ respectively. 
For each band we define two gaps,
$\Delta_{i1}$ and $\Delta_{i2}$ ($i =1,2$ is the band index), 
corresponding to these two regions.
Assuming parabolic bands as before, the gap equations are:
\be
\Delta_{11} = & - & \Delta_{11} \mu_1 \int_{0}^{\omega_{C1}} \frac{\tanh(\beta E_{11} /2)}{E_{11}} d\epsilon +  \nonumber \\
         & - &  \Delta_{21} \lambda_2 \int_{0}^{\omega_{C1}} \frac{\tanh(\beta E_{21} /2)}{E_{21}} d\epsilon -  \nonumber \\
         & - & \Delta_{12} \mu_1 \int_{\omega_{C1}}^{\omega_{C2}} \frac{\tanh(\beta E_{12} /2)}{E_{12}} d\epsilon \nonumber
\ee
\be
\Delta_{12} = & - & \Delta_{11} \mu_1 \int_{0}^{\omega_{C1}} \frac{\tanh(\beta E_{11} /2)}{E_{11}} d\epsilon -  \nonumber \\
         & - &\Delta_{12} \mu_1 \int_{\omega_{C1}}^{\omega_{C2}} \frac{\tanh(\beta E_{12} /2)}{E_{12}} d\epsilon
\label {eqGaps}
\ee
where $E_{ki} = \sqrt{\epsilon_{ki}^2 + \Delta_{ki}^2}$. 
To simplify the notation we have introduced 
new variables $\mu_{1,2} = N_{1,2} U_{1,2} $ and $\lambda_{1,2} = N_{1,2} J$,
where $N_i$ is the density-of-states (DOS) in band $i$. 
Both $\mu_i$ and $\lambda_i$ are positive corresponding
to repulsive interactions. 
The above equations are coupled with the analogous ones for $\Delta_{21}$ and $\Delta_{22}$.
The interactions $J$, $U_1$, and $U_2$ appearing  here
are the angular averages over the Fermi surface \cite{vdw}.

To proceed, we concentrate on
the condition for a non-trivial solution in terms of $\Delta_{ki}$. 
This gives an equation for the integrals $I_1$ and $I_2$ 
that appear in Eq. \eqref{eqGaps}:
\be
I_1 = \int_{0}^{\omega_{C1}} \frac{\tanh(\beta E_{11} /2)}{E_{11}} d\epsilon \nonumber\\
I_2 = \int_{\omega_{C1}}^{\omega_{C2}} \frac{\tanh(\beta E_{12} /2)}{E_{12}} d\epsilon.
\ee

These integrals are positive-definite and close to 
the critical temperature can be approximated by
$I_1 = \log{(1.13 \omega_{C1}/T_c)}$ and $I_2 = \log{(\omega_{C2}/\omega_{C1})}$. 
If we now momentarily suspended the above ''retardation'' 
effect of $J$ relative to $U_{1,2}$ ($\omega_{C2}=\omega_{C1}$)
and take the cut-off of the energy integration to be the 
same everywhere $I_2\to 0$ and $I_1$ becomes
\be 
I_1 = \frac{( \mu_1 + \mu_2)
  + \sqrt { (\mu_1 - \mu_2)^2 + 4 \lambda_1 \lambda_2}}{2 (\lambda_1 \lambda_2 - \mu_1 \mu_2)}~.
\ee

The physical solution exists {\em only} if the right-hand side is positive, which 
translates into $\lambda_1 \lambda_2 > \mu_1 \mu_2$. 
Solving for the critical temperature we get:
\be
T_c  = (1.13 \omega_C)e^{- \frac{( \mu_1 + \mu_2)
  + \sqrt { (\mu_1 - \mu_2)^2 + 4 \lambda_1 \lambda_2}}{2 ( \lambda_1 \lambda_2 - \mu_1 \mu_2)}}.
\label{eliashberg}
\ee

The full expression for $T_c$, obtained from Eq. \eqref{eqGaps},
with $\omega_{C2}>\omega_{C1}$ and under the same conditions $I_1 > 0$ and $I_2 > 0$, 
is more complicated
\be
T_c  & = & (1.13 \omega_{C1})\times \label{fulleliashberg} \\
& & e^{- \frac{( \mu_1 + \mu_2 + 2 I_2 \mu_1 \mu_2 )
  + \sqrt { (\mu_1 - \mu_2)^2 + 4 \lambda_1 \lambda_2 (1 + I_2 \mu_1)^2(1 + I_2 \mu_2)^2}}{2 (  \lambda_1 \lambda_2 (1 + I_2 \mu_1)(1 + I_2 \mu_2) - \mu_1 \mu_2)}}~. \nonumber
\ee
As it turns out, however, once the retardation effect {\em is} restored,
all the reader needs to do to obtain a good approximation to
(\ref{fulleliashberg}) is to simply
replace in \eqref{eliashberg} $\mu_{1,2}\to \mu_{1,2}^\star$,
where $\mu_{1,2}^\star$ are given by:
\be
\mu_{1,2}^\star = \frac {\mu_{1,2}} {1 + \mu_{1,2} \log({\omega_{C2}/\omega_{C1}})}~.
\ee
Therefore, now $ \lambda_1 \lambda_2$  only has to be larger than $\mu_1^\star \mu_2^\star$,
or, in the original notation $J > \sqrt{U_1^\star U_2^\star }$, to make
$s'$ superconductivity possible. Notice that, in a similar vain, any reduction in
$U_{1,2}$ arising from the interband phonon
attraction would help superconductivity as well.

One of the consequences of Eqs. \eqref{eqGaps} is that the overall magnitudes of the
hole and electron gaps, $\vert\Delta_1\vert$ and $\vert\Delta_2\vert$, are
generically different, as soon as the $c$ and $d$ band parameters are not the same.
Several point-contact Andreev reflection (PCAR) measurements \cite {CLC, Samuely}, however,
show a single gap. Obviously, one
explanation is that there are two gaps, but
they have similar magnitudes in this particular system and cannot be resolved in a
PCAR experiment.
\begin{figure}[htb]
\includegraphics[width=0.4\textwidth]{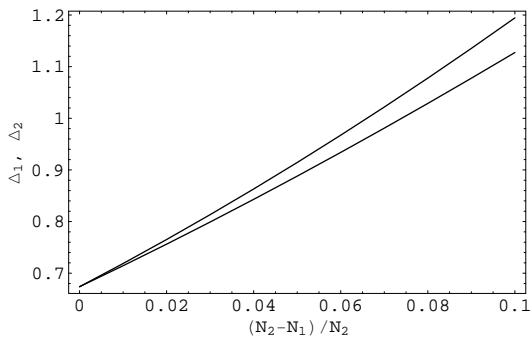}
\caption{The illustration of the relatively mild sensitivity of the two
superconducting gaps to differences in the DOSs between electron and hole
bands ($ (N_2 - N_2)/N_2$). The gaps were computed from Eq. \eqref {eqGaps}, without retardation effects,
using  $\mu_1 = 0.3$ and $\lambda_1 = 0.5$, for up to $10\%$
difference in DOSs, and are measured in units of $10^{-2} \omega_C$. }
\label {Fig_Gaps}
\end{figure}

In contrast, the ARPES experiments do in fact show two different gaps in some of the other
Fe-pnictide superconductors \cite{Ding,Hasan}.
Note that the main role of different band parameters arises through their
different densities-of-states (DOSs), the parameter which is independent
of doping for 2D parabolic bands.
Naturally, the real hole and electron bands in Fe-pnictides are neither
ideal parabolas nor identical to
each other (nor entirely 2D for that matter). To study the effects of
different DOSs for the two bands we solve
numerically Eq. \eqref{eqGaps} (without the retardation effects) at zero temperature for
differences in DOS of the two bands of up to $10\%$. The result is
shown in Fig \ref {Fig_Gaps} and we
see that the hole and electron gaps diverge away from one another rather slowly. This
illustrates the relative lack of sensitivity of the gap functions to
differences in the band parameters, retroactively provides another justification
for our assuming that they are the same, and, within the framework of the
$s'$-wave state,
can explain the experiments that seemingly point to a single-gap superconductivity.

\section {Discussion and Conclusions}

In this paper we investigated the dynamical properties of spin fluctuations in superconductors
with various forms of the microscopic order parameter, both in the vicinity of and within the
itinerant SDW state. Our main emphasis was on finding ways to distinguish different types of
superconducting order by concentrating on the features in {\em momentum} space, since one of the
leading candidates, the extended $s'$-wave superconductor, does not differ from an ordinary
$s$-wave by the overall symmetry. We demonstrated by explicit calculations that the momentum as
well as the frequency dependence of the spin fluctuation decay rate can be used to distinguish
among different states of a multiband, multigap superconducting system, even in the cases when the
temperature dependence is similar; some of our results are summarized in Table \ref{tbl:damp2}.
Among other applications, we expect our findings to be particularly useful in the ongoing efforts
to establish the symmetry of the order parameter in the iron-based high-temperature
superconductors.

\section {Acknowledgements}

We thank V. Cvetkovic, C. Broholm and W. Bao for useful discussions.
This work was supported in part
by the NSF grant DMR-0531159.

\bibliographystyle{apsrev}

\end {document}